\documentclass[journal, 10pt]{IEEEtran}

\usepackage{multirow}
\usepackage[font=scriptsize,caption=false,labelsep=space]{subfig}
\usepackage{mathrsfs}
\usepackage{graphicx,float,wrapfig,epstopdf,amsmath}
\usepackage[]{algorithmicx}
\usepackage{algpseudocode,algorithm}
\usepackage{balance}
\usepackage{slashbox}

\usepackage{mathtools,lipsum}
\usepackage{amsmath}
\usepackage{amssymb}
\usepackage{amsthm}
\usepackage{graphicx}
\usepackage{epstopdf}
\usepackage{times}
\usepackage{textcomp,cite}

\usepackage{url}
\usepackage{hyperref}

\usepackage{multirow}
\usepackage{threeparttable}
\graphicspath{{./figures/}}

\usepackage[table]{xcolor}
\definecolor{intnull}{RGB}{213,229,255}
\definecolor{inteins}{RGB}{128,179,255}
\definecolor{color1}{RGB}{199,209,232}
\definecolor{color2}{RGB}{230,231,233}

\begin{document}

	\title{ Federated Learning for Physical Layer Design
	}
	\author{\IEEEauthorblockN{Ahmet M. Elbir, Anastasios K. Papazafeiropoulos, and Symeon Chatzinotas}
		\thanks{This work was supported in part by the ERC Project AGNOSTIC.}
		\thanks{A. M. Elbir is with the Department of Electrical and Electronics Engineering, Duzce University, Duzce, Turkey, and SnT at the University of Luxembourg, Luxembourg (e-mail: ahmetmelbir@gmail.com).} 
		\thanks{A. K. Papazafeiropoulos is with the CIS Research Group, University of Hertfordshire, Hatfield, U. K., and SnT at the University of Luxembourg, Luxembourg (e-mail: tapapazaf@gmail.com). }
		\thanks{S. Chatzinotas is with the SnT at the University of Luxembourg, Luxembourg (email: symeon.chatzinotas@uni.lu). }	
	}
	\maketitle
	
	\begin{abstract}
		Model-free techniques, such as machine learning (ML), have recently attracted much interest towards the physical layer design, e.g., symbol detection, channel estimation, and beamforming. Most of these ML techniques employ centralized learning (CL) schemes and assume the availability of datasets at a parameter server (PS), demanding the transmission of data from  edge devices, such as mobile phones, to the PS. Exploiting the data generated at the edge, federated learning (FL) has been proposed recently as a distributed learning scheme, in which each device computes the model parameters and sends them to the PS for model aggregation while the datasets are kept intact at the edge. Thus, FL is more communication-efficient and privacy-preserving than CL and applicable to the wireless communication scenarios, wherein the data are generated at the edge devices. This article presents the recent advances in FL-based training for physical layer design problems. Compared to CL, the effectiveness of FL is presented in terms of communication overhead with a slight performance loss in the learning accuracy. The design challenges, such as model, data, and hardware complexity, are also discussed in detail along with possible solutions.

	\end{abstract}
	%
	

	\section{Introduction}
	\label{sec:Introduciton}
	The ever-growing increase in the number of connected devices in the last few years has led to a surge in the amount of data generated by mobile phones, connected vehicles, drones, and internet of things (IoT) devices due to the rapid development of various emerging applications, such as artificial intelligence (AI), virtual and augmented reality (VAR), autonomous vehicles (AVs), and machine-to-machine communications~\cite{fl_survey}. According to the international telecommunication union (ITU), the global mobile traffic is expected to reach $607$ EB (Exabyte) in 2025. In order to process and extract useful information from the huge amount of data, machine learning (ML) has been recognized as a promising tool for several emerging technologies, such as IoT, AVs, and the next-generation wireless communications~\cite{dl_GGui_WCM,dl_WCM}. These applications require a huge amount of data to be processed and learned by a learning model, often an artificial neural network (ANN), by extracting the features from the raw data and providing a ``meaning'' to the input via constructing a model-free data mapping with a huge number of learnable parameters set via powerful computational resources, such as graphics processing units (GPUs)~\cite{dl_WCM}. Reasonably, huge learning models, the massive amount of training data, and the powerful computation infrastructure are the main driving factors of the success of ML algorithms~\cite{dl_GGui_WCM,dl_WCM,fl_survey}.

	Most of the ML tasks are realized in a centralized manner, i.e., the learning model is trained at a central entity, e.g., a parameter server (PS). Hence, it is usually called centralized learning (CL)~\cite{fl_deniz_DSGD,elbir2020FL_HB}. The implementation of CL brings the following challenges for model training:
	\begin{itemize}
		\item The training data is usually generated/collected at the edge devices (clients), e.g., mobile phones, sensors. Thus the transmission of the training data to the PS is required so that the learning model can be trained on the collected training data. This process introduces a huge overhead because of the dataset transmission through wireless links.
		
		\item Due to the transmission of the raw datasets from the clients to the PS, the data become fragile to the attacks by third parties, thereby, introducing privacy concerns, especially when the content of the data is sensitive, e.g., images, videos \textcolor{black}{and user locations.}
		
		\item In the inference stage of CL, the clients should transmit their data to the PS for each recognition/classification task, which results in a frequent information exchange between the client and PS for the transmission of input/outputs. This is especially critical in future applications, in which most of the edge-PS data traffic is expected to be ML-related, e.g., images, videos~\cite{fl_survey,fl_spm_federatedLearning}.
	\end{itemize}
	
\begin{figure*}[h]
		\centering
		{\includegraphics[draft=false,width=.8\textwidth]{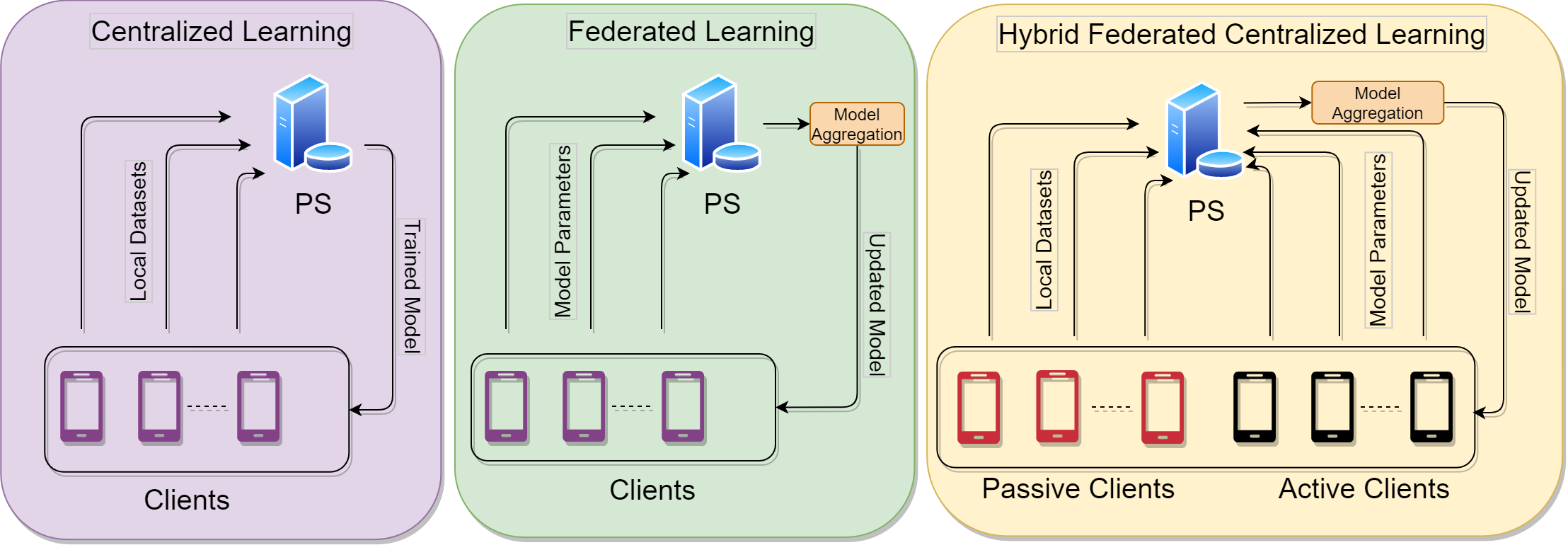} }
		\caption{ \textcolor{black}{CL (Left), FL (Middle), and HFCL (Right)} frameworks for model training. While the clients send their local datasets to the PS in CL, FL involves the transmission of only the model parameters during training. In the hybrid scenario (HFCL), a portion of the clients (passive) sends their local datasets while the remaining ones (active) send the model parameters to the PS, which aggregates them and broadcasts to the clients. }
		\label{fig_CL_FL}
	\end{figure*}

	To answer the aforementioned challenges, distributed ML strategies, such as federated learning (FL), have been envisioned by \textcolor{black}{McMahan \emph{et al.}}~\cite{fl_By_Google} to bring the learning task to the edge level instead of training the model at a central entity, as illustrated in Figure~\ref{fig_CL_FL}.	In FL, instead of sending their local datasets, the clients compute the model parameters and send them to a central entity, i.e., the PS, which aggregates the model parameters and broadcasts them to the clients (see, e.g., Figure~\ref{fig_CL_FL}). As a result, the computation of the model parameters is handled at the edge level. Thus, the clients can, then, use the trained model for recognition/classification tasks without sending the raw data to the PS. \textcolor{black}{Although FL requires many communication rounds between the PS and the clients for model convergence, compared to CL, FL enjoys a significant reduction in terms of the communication overhead at the cost of a slight performance loss due to the absence of access to the whole data at once as in CL. Furthermore, FL is more privacy-preserving since it does not involve the transmission of datasets~\cite{fl_By_Google,fl_deniz_DSGD}.} Motivated by these advantages, FL has gained significant interest recently, especially for the computer vision (CV) tasks, such as image classification.

	With the promising performance of ML in CV tasks~\cite{fl_survey}, there has been a surge in the amount of ML-based works on the physical layer design for wireless communication systems~\cite{dl_phy_GeoffreyLi1,dl_GGui_WCM,dl_WCM}, while most of these works have utilized CL-based schemes. However, the application of FL  fits perfectly to the physical layer design scenario since the data, e.g., the received symbols, the channel state information (CSI), and the beamformers weight, are generated at the edge level. For instance, the received pilots, channel response, and beamformer information can be used to generate datasets for symbol detection~\cite{fl_symbolDetection}, channel estimation~\cite{elbir2020_FL_CE}, and beamformer design~\cite{elbir2020FL_HB}, respectively. Therefore, the employment of FL can be more appropriate compared to CL for physical layer design. In such a scenario, each edge client, e.g., the mobile phones, can compute the model parameters based on their local datasets and the model aggregation can be performed at the PS by sending the model updates to the PS through the base stations (BSs). As a result, it is expected to witness a significant amount of works on the FL-based counterparts of the aforementioned CL-based schemes in the future.


	This article provides a comprehensive analysis of the FL-based learning models for physical layer design of communication systems in terms of training data collection, model training, and deployment for inference purposes. \textcolor{black}{While an overview of FL has been provided in the relevant literature, e.g.,~\cite{fl_spm_federatedLearning,fl_survey6G,fl_dist_survey}, they either consider non-physical layer applications, such as image/speech/text recognition/classification~\cite{fl_spm_federatedLearning}, or a general FL framework~\cite{fl_survey6G,fl_dist_survey}. In contrast, this work focuses on the physical layer design and provides a synopsis of various FL-based applications, such as symbol detection~\cite{fl_symbolDetection}, channel estimation~\cite{elbir2020_FL_CE}, and beamforming~\cite{elbir2020FL_HB,elbir2021ICASSP_FL_HB_SPIM,fl_IRS_Beamforming_RateOpt}, which are the main FL-based applications in the literature.} In the remainder of this article, we first briefly discuss how FL works and what distinguishes it from CL along with its advantages and drawbacks. Following that, FL-based solutions for the physical layer design are provided. We then identify the design and research challenges in FL, such as data non-uniformity, model and data complexity, hardware requirements, etc, and highlight possible future research directions that may be helpful in imparting FL to the physical layer design to achieve more data-, model- and hardware-efficiency.

	\section{Federated Learning versus Centralized Learning}
	
	\subsection{Centralized Learning}
	In CL, the clients send their local datasets, as shown in Figure~\ref{fig_CL_FL}, to the PS for training, which can be done via gradient descent (GD) or stochastic GD (SGD) techniques~\cite{fl_deniz_DSGD,fl_By_Google}. The main advantage of CL is the access to the whole training dataset at once so that more accurate learning performance can be achieved. However, the main drawback of CL is the requirement of dataset transmission, which entails huge communication overhead, as shown in Table~\ref{tableComparison}.

	\subsection{Federated Learning}
	FL-based training has three stages. In the first stage, training data is collected in an offline manner. The second phase involves model computation at the edge and aggregation at the PS. Finally, in the third phase, the trained model is deployed for inference. 
	In contrast to CL, FL involves only the transmission of either model parameters~\cite{fl_By_Google,elbir2020_FL_CE} or the model updates (gradients)~\cite{elbir2020FL_HB,fl_symbolDetection} between the clients and the PS, as shown in Figure~\ref{fig_CL_FL}. While the transmission of model parameters directly provides a better representation of the learning model, a gradient-based transmission is more energy-efficient due to the requirement of lower transmit power~\cite{fl_deniz_DSGD}. Thus, the FL still requires a central entity, whose only purpose is to aggregate the models and broadcast them.	\textcolor{black}{Therefore, FL does not involve dataset transmission as in CL. As a result, FL is more communication-efficient since the size of the training dataset is usually larger than the size of the model parameters in ML tasks~\cite{fl_By_Google,fl_deniz_DSGD,elbir2020FL_HB}.}
	

	\begin{table}[t]
		\caption{Comparison of CL, FL and HFCL Frameworks}
		\label{tableComparison}
		\centering
		\begin{tabular}{c|c|c|c }
			\hline
			\hline
			\backslashbox{\textbf{Property}}{\bf Framework}\cellcolor{color1} &\bf CL  
			\cellcolor{color2} 
			&\bf FL
			\cellcolor{color1} &\bf HFCL\cellcolor{color2} \\
			\textbf{Communication Overhead}\cellcolor{color2} & High\cellcolor{color1} & Low\cellcolor{color2} &Moderate\cellcolor{color1} \\
			\hline
			\textbf{Learning Accuracy}\cellcolor{color1} & High\cellcolor{color2} & Moderate\cellcolor{color1} &Moderate\cellcolor{color2} \\
			\hline
			\hspace{-3pt}\textbf{Hardware Requirement at The Edge}\hspace{-3pt}\cellcolor{color2} & \cellcolor{color1}Low& High\cellcolor{color2} &Flexible \cellcolor{color1} \\
			\hline
			\textbf{Privacy-Preserving}\cellcolor{color1} & Low\cellcolor{color2} & High\cellcolor{color1} &Moderate\cellcolor{color2} \\
			\hline
			\hline
		\end{tabular}
	\end{table}

	\subsection{Hybrid Federated and Centralized Learning}
	Bringing the learning task to the edge level demands computational capability from the clients for model computation usually by using GPUs, which may not always be available at the edge devices. To solve this issue, hybrid federated and CL (HFCL) techniques can be employed, as illustrated in Figure~\ref{fig_CL_FL}. The main idea behind HFCL is that all of the clients can participate in training regardless of their computational capability. Thus, the clients (\textit{active}) which have enough computational resources perform FL while the remaining ones (\textit{passive}), which lack such resources, resort to CL~\cite{elbir2021HFCL}. As a result, we can benefit from accessing the dataset of passive clients, which would not be accessed if only FL was employed. Thus, HFCL provides a trade-off between the higher learning accuracy of CL and the reduced communication overhead of FL (Table~\ref{tableComparison}).

	%
	%
	%

	\section{Federated Learning in the Physical Layer}
	In this section, we present  FL applications in the physical layer design, such as symbol detection, channel estimation, and beamforming, as illustrated in Figure~\ref{fig_FL_PHY}. The summary of the benefits and drawbacks of FL-based techniques is provided in Table~\ref{tableSummary}. 
	
	\textcolor{black}{The input-output pair for symbol detection is received-transmitted symbols~\cite{fl_symbolDetection}, while the learning model in channel estimation accepts the input as the received pilot signals and the output is the channel matrix~\cite{elbir2020_FL_CE}. In beamforming, the learning model can be either fed with the received pilots or channel matrix and the output layer can be designed via either regression (beamformer weights)~\cite{fl_IRS_Beamforming_RateOpt,elbir2021ICASSP_FL_HB_SPIM} or classification (beamformer indices) layers~\cite{elbir2020FL_HB}.	The selection of the learning model is also critical. Specifically, multilayer perceptrons (MLPs) employed in~\cite{fl_symbolDetection} and \cite{fl_IRS_Beamforming_RateOpt} have only fully connected layers, which are useful for data mapping. However, the convolutional neural networks (CNNs) benefit from using both fully connected and convolutional layers, which are powerful for feature extraction from the input data~\cite{elbir2020_FL_CE,elbir2021ICASSP_FL_HB_SPIM}.}
	
	\subsection{Symbol Detection}
\begin{figure}[t]
		\centering
		{\includegraphics[draft=false,width=\columnwidth]{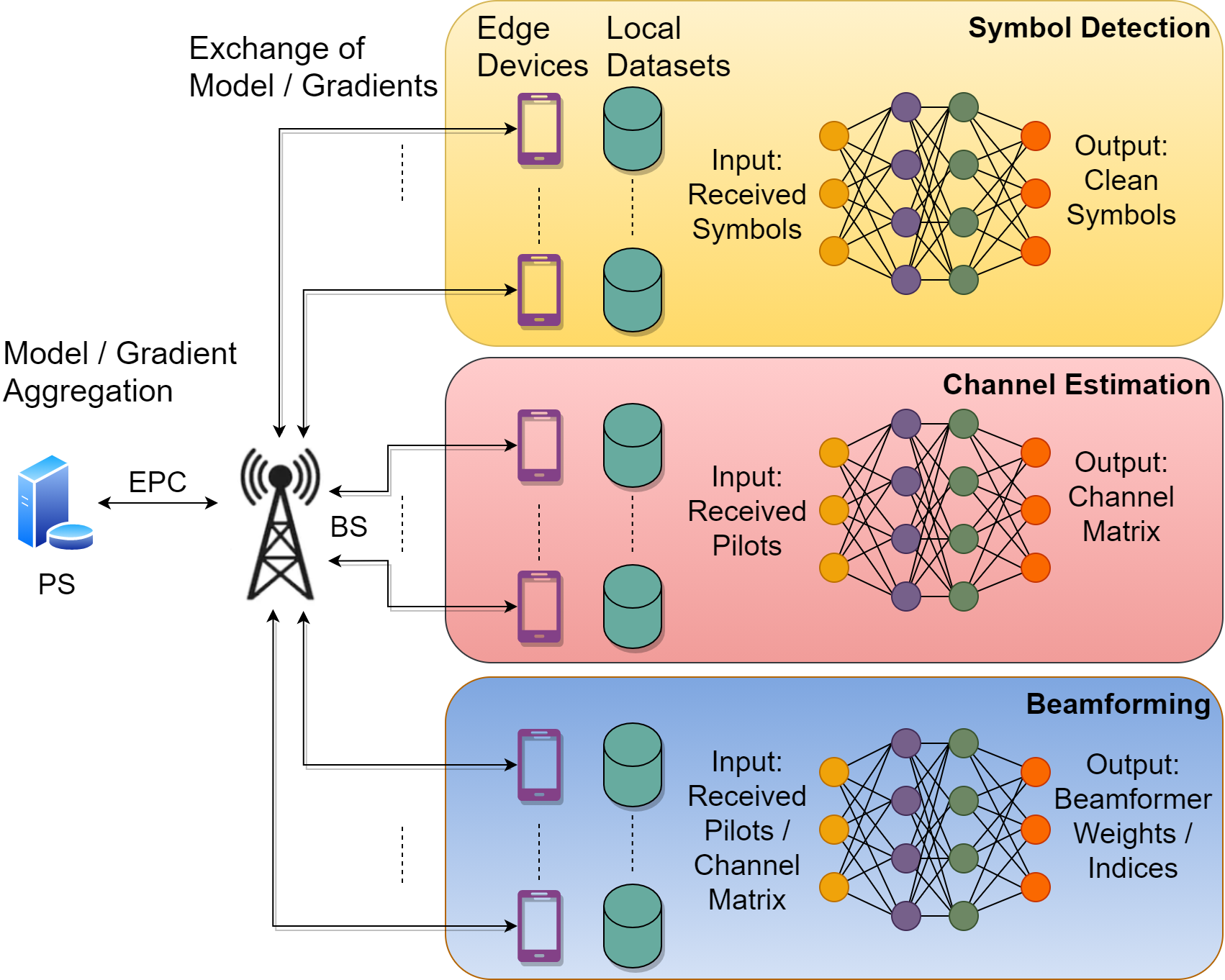} }
		\caption{FL-based physical layer design applications, such as symbol detection, channel estimation, and beamforming. The learning model parameters computed at the edge devices are sent to the PS through the BS for model aggregation. }
		\label{fig_FL_PHY}
	\end{figure}

	Symbol detection via ML provides an end-to-end learning scheme such that the received symbols under the effect of wireless channel are mapped to the clean symbols via ANNs~\cite{dl_WCM,dl_phy_GeoffreyLi1} (see Figure~\ref{fig_FL_PHY}). The main advantage of using ML for symbol detection is to provide a data-driving mapping for modeling the channel characteristics, which may not be accurately handled via model-based techniques~\cite{fl_symbolDetection}. Furthermore, end-to-end learning enables the model to detect the symbols accurately without the channel estimation stage because it is directly fed with the received corrupted symbols~\cite{dl_phy_GeoffreyLi1}. Employing CL for this task raises challenges in terms of communication overhead as well as privacy concerns such that third parties can detect and cache the data during dataset transmission~\cite{fl_spm_federatedLearning}. 
	
	The FL-based receiver (\texttt{FedRec}) is proposed by \textcolor{black}{Mashhadi \emph{et al.}}~\cite{fl_symbolDetection} for symbol detection in downlink fading channels. Specifically, a gradient-based federated architecture is considered and the collected gradient information is sent to the PS via the evolved packet core (EPC)~\cite{fl_survey}. The main advantage of this approach is the substantial reduction in the communication overhead while maintaining a satisfactory detection performance. However, \cite{fl_symbolDetection} employs a very shallow MLP structure with a single hidden layer and only $48$ learnable parameters, which makes the learning performance strongly dependent on the input-output data space, thus wider and deeper model architectures are required for the generalization. Furthermore, generative adversarial networks (GANs) can be employed for data enrichment if the dataset at hand is insufficient. A GAN is composed of two neural networks: the generator and the discriminator. The discriminator attempts to differentiate between the real and fake data while the latter is produced by the generator to fool the discriminator. Thus,  GAN learns to generate new data with the same statistics as the training set~\cite{dl_phy_GeoffreyLi1,fl_dist_survey}.

	\begin{table*}[t]
		\caption{FL-based Techniques for Physical Layer Design
		}
		\label{tableSummary}
		\centering
		\begin{tabular}{p{0.18\textwidth}p{0.22\textwidth}p{0.22\textwidth}p{0.22\textwidth}}
			\hline 
			\hline
			\bf Application \cellcolor{color1}
			&\cellcolor{color2}\bf Learning Scheme
			& \bf Benefits\cellcolor{color1}
			&\cellcolor{color2}\bf Drawbacks 
			\\
			\hline
			{ Signal Detection} \cite{fl_symbolDetection}\cellcolor{color1}
			&\cellcolor{color2}	 Gradient transmission for a regression MLP model
			& \cellcolor{color1} No need for channel estimation algorithm
			&Deeper models are required to generalize the learning performance \cellcolor{color2} 
			\\
			\hline
			{ Channel Estimation} \cite{elbir2020_FL_CE}\cellcolor{color1}
			&	\cellcolor{color2}Model transmission for a CNN predicting the direct and cascaded IRS channels
			&\cellcolor{color1} Each user possesses the trained model and can estimate its own channel with it
			&\cellcolor{color2} Data collection requires a labeling process to obtain the channel data
			\\
			\hline
			\cellcolor{color1}
			{ Beamforming in Massive MIMO} \cite{elbir2020FL_HB}
			&\cellcolor{color2} Gradient transmission for a classification CNN model
			&\cellcolor{color1} Simple beamformer construction from the predicted indices at the classification layer
			&\cellcolor{color2} Sub-optimum performance due to codebook-based beamformer design
			\\
			\hline
			\cellcolor{color1}
			{ Beamforming With SPIM} \cite{elbir2021ICASSP_FL_HB_SPIM}
			&\cellcolor{color2}	Gradient transmission for a regression CNN model with dropout layers
			&\cellcolor{color1} Higher spectral efficiency can leverage the performance loss due to distributed training
			&\cellcolor{color2} Labeling requires heavy computation resources
			\\
			\hline
			\cellcolor{color1}
			{ Beamforming in IRS-aided Massive MIMO} \cite{fl_IRS_Beamforming_RateOpt} 
			&\cellcolor{color2}	 Model transmission for a regression MLP model 
			&\cellcolor{color1} Accelerated convergence thanks to local model updates
			&\cellcolor{color2} Only IRS-beamforming is performed
			\\
			\hline
			\hline 
		\end{tabular}
	\end{table*}

	\subsection{Channel Estimation}
	Via conventional ML, several channel estimation methods have been introduced recently~\cite{dl_phy_GeoffreyLi1,dl_WCM,dl_GGui_WCM}, while all of these works assume the availability of a central entity, at which the learning model is trained by collecting the channel matrix either at the users (downlink) or at the BS (uplink).	
	
\begin{figure}[t]
		\centering
		{\includegraphics[draft=false,width=\columnwidth]{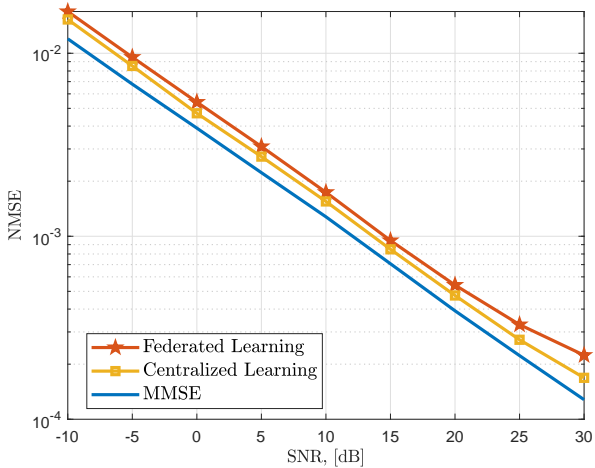} } 
		\caption{Channel estimation NMSE versus SNR in IRS-aided massive MIMO.
		}
		\label{fig_CE_SNR}
	\end{figure}

	FL-based channel estimation has been considered in~\cite{elbir2020_FL_CE} for the downlink scenario of both conventional and intelligent reflecting surface (IRS)-aided millimeter-wave (mm-Wave) massive multiple-input multiple-output (MIMO) systems. A CNN is utilized via an FL-based approach in three phases. First, the training data is collected, in which the received pilot signals (input) and the channel matrix (output) are obtained, where a model-based approach, such as approximate linear minimum mean-squared-error (A-LMMSE) estimation, is used to obtain the labels. In the second phase, the users collaboratively train a model via FL by computing the model parameters and exchange them with the PS. The third phase is the prediction phase, where the trained model becomes available to each user that can feed it with the received pilots to predict the corresponding downlink channel. The channel estimation performance of FL and CL together with MMSE is depicted in Figure~\ref{fig_CE_SNR} in terms of normalized MSE (NMSE) with respect to the signal-to-noise-ratio (SNR) for an IRS-aided scenario when there are $8$ users and the number of antennas at the BS and the IRS are $128$ and $100$, respectively. We can see a slight performance loss of FL over CL, while the former achieves approximately $10$-fold reduction in terms of the communication overhead than the latter~\cite{elbir2020_FL_CE}. A major advantage of this approach is that each user has access to the trained model to perform channel estimation. However, it involves a training data generation phase, in which the labels (channel matrix) of the training dataset should be obtained via another model-based technique. Thus, the performance of the learning model becomes upper-bounded by this benchmark technique~\cite{dl_phy_GeoffreyLi1}. Label-free approaches, such as reinforcement learning (RL) can be integrated into FL so that instead of obtaining the channel data from a model-based approach, a cost function can be properly defined and optimized by RL to predict the channel data.

	\subsection{Beamforming}
	While symbol detection and channel estimation are the core subjects in the field of wireless communications, beamforming has been a vital part of it, recently, due to the advances in multi-antenna communications employing very large antenna arrays to benefit from the high beamforming gain. \textcolor{black}{This property can compensate for the high path loss in high-frequency bands (mm-Wave to terahertz (THz)).} As a result, massive MIMO has been envisioned in the deployment of the fifth-generation (5G) systems and it is expected to be one of the main pillars of beyond 5G (B5G) networks~\cite{dl_WCM,fl_survey6G}. To reduce the hardware complexity and cost, massive MIMO in 5G has employed hybrid analog-digital beamformers so that only a small number of radio frequency (RF) chains can be used. 
	
	\subsubsection{Beamforming in Massive MIMO} FL-based hybrid beamforming is proposed in~\cite{elbir2020FL_HB} for the downlink scenario, wherein the users compute the model parameters and send the gradient information to the BS. A classification CNN model is trained such that each class designates a possible user direction in the azimuth plane and the analog beamformers are constructed as the steering vectors corresponding to these user directions. As a result, the output of the CNN is the beamformer indices. The input of the CNN can be either the channel matrix or the received pilot signals that are used as input in the channel estimation case. The main trade-off between these two inputs is that the former provides more features than the latter while the latter is easy to collect rather than performing channel estimation to obtain the former~\cite{dl_GGui_WCM,elbir2020_FL_CE}.

\begin{figure}[t]
		\centering
		{\includegraphics[draft=false,width=\columnwidth]{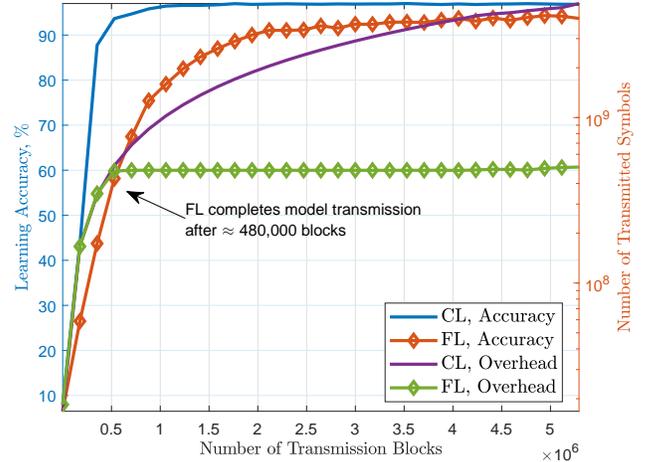} }
		\caption{\color{black}Communication Overhead and learning accuracy of CL and FL in hybrid beamformer design~\cite{elbir2020FL_HB}. 
		}
		\label{fig_FL_overhead}
	\end{figure}

	\textcolor{black}{Figure~\ref{fig_FL_overhead} shows the communication overhead and learning accuracy for the hybrid beamforming problem for FL and CL when the learning model (FL) dataset (CL) is transmitted to the PS for $1000$ symbols per transmission block~\cite{elbir2020FL_HB}. It can be seen that FL completes the model parameter transmission $10$ times quicker than CL, in which the local datasets are transmitted. The communication efficiency of FL is at the cost of a slight performance loss in the learning accuracy since it does not have access to the whole dataset at once. }

	To improve the spectral efficiency in mm-Wave systems, index modulation (IM) is an attractive technique, in which additional information is encoded in the indices of the transmission media such as, subcarriers, antennas, or spatial paths. FL-based beamforming with spatial path index modulation (SPIM) is considered in~\cite{elbir2021ICASSP_FL_HB_SPIM}, where a regression CNN model with dropout layers (DLs) is trained. DLs are useful to mitigate overfitting and for the generalization of the learning model. In a DL with a $50\%$ rate, only half of its weights are updated in each iteration of training. Thus, the use of DL is significant for communication efficiency. It is reported in~\cite{elbir2021ICASSP_FL_HB_SPIM} that FL with (without) DL enjoys approximately $10$	($5$) times lower overhead than CL. 
	
	The communication efficiency can be further improved if sparsification techniques are applied to the model parameters or to the  gradient information that are transmitted during training~\cite{fl_symbolDetection,fl_deniz_DSGD}. This can be done by selecting a portion of the most significant of the model parameters and discarding the rest. At the beginning of the training, the deviations in the model parameters are large and the data may not be sparse. However, as the model converges, the changes in the model parameters become small so that sparsification can be extremely helpful and allow to ignore the incremental changes.
	
	The major drawback of~\cite{elbir2021ICASSP_FL_HB_SPIM} is the labeling stage, which is conducted via alternating manifold optimization (MO) techniques, which may require heavy computational resources, especially for a high number of antennas and RF chains, which is the case in massive MIMO systems. A codebook-based approach can be followed to reduce the cost of labeling while sacrificing the spectral efficiency ~\cite{elbir2020FL_HB}.

	\subsubsection{Beamforming in IRS-assisted Massive MIMO}Compared to conventional massive MIMO, the IRS-aided scenario provides several advantages~\cite{fl_IRS_Beamforming_RateOpt,elbir2020_FL_CE}. IRS-aided massive MIMO presents a more green and suitable solution to enhance the wireless network performance with low cost and reduced complexity due to the employment of passive elements in the IRS. An IRS is an electromagnetic 2-D surface that is composed of a large number of passive reconfigurable meta-materials, which reflect the incoming signal from the BS towards the users. Hence, the usage of IRS improves the received signal energy at the distant users as well as expanding the coverage of the BS. 
	
	Along with the aforementioned advantages, IRS-aided massive MIMO involves a more complex design problem than conventional massive MIMO. For instance, while the latter includes a single channel link between the BS and the user, the former involves multiple communication links, i.e., BS-user (direct) and BS-IRS-user (cascaded) links~\cite{elbir2020_FL_CE,fl_survey6G}. Even more challenging, an IRS-aided scenario involves both active (BS) and passive (IRS) beamforming so that the received signal from the BS can be reflected to the user. In addition, the optimization of the beamformers is based on the channel links, which makes the IRS-aided case very challenging, and thus several approaches have been proposed to overcome these difficulties~\cite{fl_IRS_Beamforming_RateOpt,elbir2020_FL_CE}.

	%
	
	FL-based beamformer design for IRS-aided scenario has been proposed by \textcolor{black}{Ma \emph{et al.}}~\cite{fl_IRS_Beamforming_RateOpt}, where the \texttt{FedAvg} approach~\cite{fl_By_Google} is adopted to train a regression MLP with model transmission. In \texttt{FedAvg}, the model parameters computed at the edge devices are averaged for a certain number of iterations. Then, they are sent to the PS, where a global aggregation is performed, thereby, improving the convergence rate~\cite{fl_By_Google,fl_deniz_DSGD}. In~\cite{fl_IRS_Beamforming_RateOpt}, only the channel corresponding to the IRS-user link is designed while the BS-IRS link is assumed to be constant. However, in a realistic scenario, the mm-Wave channel is very dynamic due to the environmental changes and has a very short channel coherence interval~\cite{elbir2020FL_HB,elbir2020_FL_CE}. Thus, the performance of the trained model cannot reflect the realistic propagation environment for such IRS-aided scenarios.


	%
	%

	\section{Design Challenges and Future Research Directions}
	\subsection{Non-uniformity in the Physical Layer Data}
	Since the clients collect local datasets independent of each other, there is an unavoidable non-uniformity in the whole dataset due to non-stationarity and location-dependence of the wireless channels, which may also have different models from client to client. As a result, the convergence may be longer and the global consensus among the clients may not be reached~\cite{fl_deniz_DSGD,elbir2020FL_HB}. 

	A possible solution to deal with the non-uniformity and obtain a satisfactory learning performance is to build wider and deeper models so that location-dependent data features can be distinguished and a better representation can be achieved. Thus, instead of simple MLP architectures, deeper CNNs specifically tailored for physical layer problems with a large number of convolutional layers should be employed to learn better the various features in the data. To accelerate the convergence of these deeper models, faster learning strategies, such as zero-shot learning, can be employed. In addition, hierarchical FL (HFL) schemes can be employed to cluster the clients that have close uniformity, then train multiple models for each cluster and aggregate the resulting models~\cite{fl_deniz_DSGD,fl_survey}.
	
	Another solution could be that the clients send a portion of their data to the PS. Thus, an inference on the collected dataset can be performed	to reduce non-uniformity. In HFCL~\cite{elbir2021HFCL}, only a portion of the clients sends their data to the PS. This approach can be modified such that each client only sends a portion of its dataset to the PS, which can compute model parameters on this dataset and aggregate the resulting model with the ones collected from the clients. 
	
	\textcolor{black}{
		\subsection{Data Collection and Complexity}
		Collecting physical layer data, e.g., the channel matrix coefficients, beamformer weights, and complex symbol data is a vital task requiring a good coverage of the whole data space so that the learning model can perform accurately. Due to its distributed architecture, FL has the advantage of collecting the enriched data from the distributed edge devices. Furthermore, FL provides easier access to the data generated at the edge as compared to CL. Most of the FL models in~\cite{fl_symbolDetection,elbir2020_FL_CE,elbir2020FL_HB,elbir2021ICASSP_FL_HB_SPIM,fl_IRS_Beamforming_RateOpt} are trained on relatively small datasets, which sample only a portion of the whole data space. Therefore, it is required to build huge datasets for reliable inference.} 
	
	\textcolor{black}{For very large datasets, model computation via GD or SGD becomes a tedious task for complex datasets. One way of dealing with data complexity is to select the input features accordingly.} For example, instead of feeding the learning model with channel matrix as in~\cite{elbir2020FL_HB,elbir2021ICASSP_FL_HB_SPIM} for beamforming tasks, location-only data can be used. Compared to the whole channel matrix, location information can be prepared in smaller sizes, thereby, the training stage can be accelerated.
	
	Another approach is to pre-process the input data via feature extraction techniques to provide better representations. In this way, only the necessary/important features can be extracted for inference tasks. Approaches like principal component analysis (PCA) or autoencoders can be very helpful. For instance, instead of selecting the received pilot signals as input in the channel estimation task, eigenvectors (singular value matrix) corresponding to a few largest eigenvalues (singular values) of the input data can be selected.

	\subsection{Model Complexity}
	To better extract features from complex datasets, wider and deeper models are needed.
	Training these huge models with FL can be prohibitively complex and demands huge bandwidth resources to transmit the learnable parameters.	A common solution is to prune the model parameters based on their significance. In~\cite{fl_deniz_DSGD}, a large portion of the learnable parameters that are below a certain threshold are equalized to zero and only a small portion of them are transmitted to the PS. It is reported by \textcolor{black}{Amiri \emph{et al.}}~\cite{fl_deniz_DSGD} that pruning can enable us to have satisfactory learning performance while a $50\%$ ($60\%$) reduction in the number of model parameters (transmit power) can be achieved for image classification tasks. Quantization of the learning parameters also helps to reduce the communication overhead since up to $20$-fold improvement compared to CL can be achieved for beamformer design problem as shown in~\cite{elbir2020FL_HB}. Motivated by these works, new pruning, sparsification, and quantization techniques can be developed for communication and energy-efficient FL-based training.

	\subsection{Hardware Complexity}
	A major assumption made in FL-based training is that the edge devices have sufficient computational resources, such as GPUs, to perform model computation. However, this issue is mostly overlooked in the literature~\cite{fl_By_Google,fl_survey,fl_spm_federatedLearning,fl_deniz_DSGD}.	Especially, in the presence of complex models/datasets, this assumption cannot be satisfied. The hybrid framework, in which only the clients, having sufficient computational capability, perform FL, can be a solution~\cite{elbir2021HFCL}.	It is reported in~\cite{elbir2021HFCL} that HFCL can provide approximately a $10\%$ ($20\%$) performance improvement in learning accuracy compared to conventional FL with all (only active) clients. The effectiveness of HFCL is attributed to the fact that it provides access to the data of all clients whereas FL with only active clients can use only the dataset of active clients.

	\section{Summary and Concluding Remarks}
	This article provides a synopsis of the FL-based techniques for the design of the physical layer in wireless communication systems. We have extensively investigated the existing FL-based schemes in major applications, such as symbol detection, channel estimation, and beamforming, along with their advantages and drawbacks in terms of model/data/hardware complexity, communication efficiency, and learning accuracy. The major challenges have been identified and possible solutions have been provided. We have concluded the following remarks:
	\begin{itemize}
		\item Labeling is computationally prohibitive in training data generation. Label-free learning techniques, such as RL or unsupervised learning (UL) are encouraged to reduce this complexity.
		
		\item FL demands huge computational resources from the clients, which cannot always be satisfied due to the diversity of devices with various hardware constraints. HFCL-based techniques are helpful to overcome this issue while sacrificing the privacy of a portion of the clients' data.
		
		\item Deeper models are required to generalize the learning performance and adapt to the propagation environment. This is especially demanding to cope with the complexity and non-uniformity of the physical layer data.
		
		\item Model pruning, sparsification, and quantization operations are helpful to further reduce the communication overhead in FL. In addition, feature extraction techniques can be employed to provide better data representation and lower the size of the input data to feed the model.

	\end{itemize}

	\bibliographystyle{IEEEtran}
	\bibliography{IEEEabrv,references_086}

\begin{thebibliography}{10}
\providecommand{\url}[1]{#1}
\csname url@samestyle\endcsname
\providecommand{\newblock}{\relax}
\providecommand{\bibinfo}[2]{#2}
\providecommand{\BIBentrySTDinterwordspacing}{\spaceskip=0pt\relax}
\providecommand{\BIBentryALTinterwordstretchfactor}{4}
\providecommand{\BIBentryALTinterwordspacing}{\spaceskip=\fontdimen2\font plus
\BIBentryALTinterwordstretchfactor\fontdimen3\font minus
  \fontdimen4\font\relax}
\providecommand{\BIBforeignlanguage}[2]{{%
\expandafter\ifx\csname l@#1\endcsname\relax
\typeout{** WARNING: IEEEtran.bst: No hyphenation pattern has been}%
\typeout{** loaded for the language `#1'. Using the pattern for}%
\typeout{** the default language instead.}%
\else
\language=\csname l@#1\endcsname
\fi
#2}}
\providecommand{\BIBdecl}{\relax}
\BIBdecl

\bibitem{fl_survey}
W.~Y.~B. Lim \emph{et~al.}, ``{Federated Learning in Mobile Edge Networks: A
  Comprehensive Survey},'' \emph{IEEE Commun. Surv. Tutorials}, vol.~22, no.~3,
  pp. 2031--2063, 2020.

\bibitem{dl_GGui_WCM}
H.~{Huang} \emph{et~al.}, ``{Deep Learning for Physical-Layer 5G Wireless
  Techniques: Opportunities, Challenges and Solutions},'' \emph{{IEEE} Wireless
  Commun.}, vol.~27, no.~1, pp. 214--222, 2020.

\bibitem{dl_WCM}
L.~{Dai} \emph{et~al.}, ``{Deep Learning for Wireless Communications: An
  Emerging Interdisciplinary Paradigm},'' \emph{{IEEE} Wireless Commun.},
  vol.~27, no.~4, pp. 133--139, 2020.

\bibitem{fl_deniz_DSGD}
M.~M. Amiri \emph{et~al.}, ``{Machine Learning at the Wireless Edge:
  Distributed Stochastic Gradient Descent Over-the-Air},'' \emph{IEEE Trans.
  Signal Process.}, vol.~68, pp. 2155--2169, 2020.

\bibitem{elbir2020FL_HB}
A.~M. Elbir \emph{et~al.}, ``{Federated Learning for Hybrid Beamforming in
  mm-Wave Massive MIMO},'' \emph{IEEE Commun. Lett.}, vol.~24, no.~12, pp.
  2795--2799, Aug 2020.

\bibitem{fl_spm_federatedLearning}
T.~{Li} \emph{et~al.}, ``{Federated Learning: Challenges, Methods, and Future
  Directions},'' \emph{{IEEE} Signal Process. Mag.}, vol.~37, no.~3, pp.
  50--60, 2020.

\bibitem{fl_By_Google}
B.~McMahan \emph{et~al.}, ``{Communication-Efficient Learning of Deep Networks
  from Decentralized Data},'' in \emph{International Conference on Artificial
  Intelligence and Statistics}, vol.~54, Apr 2017, pp. 1273--1282.

\bibitem{dl_phy_GeoffreyLi1}
Z.~{Qin} \emph{et~al.}, ``{Deep Learning in Physical Layer Communications},''
  \emph{{IEEE} Wireless Commun.}, vol.~26, no.~2, pp. 93--99, 2019.

\bibitem{fl_symbolDetection}
\BIBentryALTinterwordspacing
M.~B. Mashhadi \emph{et~al.}, ``{FedRec: Federated Learning of Universal
  Receivers over Fading Channels},'' \emph{arXiv}, Nov 2020. [Online].
  Available: \url{https://arxiv.org/abs/2011.07271v1}
\BIBentrySTDinterwordspacing

\bibitem{elbir2020_FL_CE}
A.~M. {Elbir} \emph{et~al.}, ``{Federated Learning for Channel Estimation in
  Conventional and IRS-Assisted Massive MIMO},'' \emph{arXiv preprint
  arXiv:2008.10846}, 2020.

\bibitem{fl_survey6G}
\BIBentryALTinterwordspacing
Z.~Yang \emph{et~al.}, ``{Federated Learning for 6G: Applications, Challenges,
  and Opportunities},'' \emph{arXiv}, Jan 2021. [Online]. Available:
  \url{https://arxiv.org/abs/2101.01338v1}
\BIBentrySTDinterwordspacing

\bibitem{fl_dist_survey}
\BIBentryALTinterwordspacing
M.~Chen \emph{et~al.}, ``{Distributed Learning in Wireless Networks: Recent
  Progress and Future Challenges},'' \emph{arXiv}, Apr 2021. [Online].
  Available: \url{https://arxiv.org/abs/2104.02151v1}
\BIBentrySTDinterwordspacing

\bibitem{elbir2021ICASSP_FL_HB_SPIM}
A.~M. Elbir \emph{et~al.}, ``Federated dropout learning for hybrid beamforming
  with spatial path index modulation in multi-user mmwave-mimo systems,'' in
  \emph{2021 IEEE International Conference on Acoustics, Speech and Signal
  Processing (ICASSP)}, 2021, pp. 8213--8217.

\bibitem{fl_IRS_Beamforming_RateOpt}
D.~Ma \emph{et~al.}, ``{Distributed Rate Optimization for Intelligent
  Reflecting Surface with Federated Learning},'' \emph{2020 ICC Workshops}, pp.
  1--6, Jul 2020.

\bibitem{elbir2021HFCL}
A.~M. Elbir \emph{et~al.}, ``{Hybrid Federated and Centralized Learning},'' in
  \emph{{29th European Signal Processing Conference (EUSIPCO)}}, 2021, pp.
  1--5.

\end{thebibliography}

	\begin{IEEEbiographynophoto} {Ahmet M. Elbir} (Senior Member, IEEE) is currently a Research Fellow at Duzce University and the University of Luxembourg. 
	\end{IEEEbiographynophoto}

	\begin{IEEEbiographynophoto} {Anastasios K. Papazafeiropoulos} (Senior Member, IEEE) is currently a Vice-Chancellor Fellow at the University of Hertfordshire and a Research Fellow at the University of Luxembourg. 
	\end{IEEEbiographynophoto}

	\begin{IEEEbiographynophoto} {Symeon Chatzinotas} (Senior Member, IEEE) is currently the Co-Head of the research group SIGCOM in the Interdisciplinary Centre for Security, Reliability and Trust, University of Luxembourg.
	\end{IEEEbiographynophoto}
	%

\end{document}